\documentclass[twocolumn,prl,aps,showpacs]{revtex4}

\usepackage{graphicx}

\begin{document}

\newcommand{\ie}{\textit{i.e.}}
\newcommand{\etal}{\textit{et al.}}
\newcommand{\eg}{e.g.}
\newcommand{\hc}{{\mathrm{h.c.}}}
\newcommand{\chem}[1]{{$\mathrm{#1}$}}
\newcommand{\Eq}[1]{Eq.~(\ref{#1})}

\title{Unconventional Superconductivity in MgCNi$_3$}

\author{Klaus Voelker}

\author{Manfred Sigrist}

\affiliation{Theoretische Physik, ETH H\"onggerberg,
             CH-8093 Z\"urich, Switzerland}

\date{August 19, 2002}

\begin{abstract}
\noindent
A multiband superconductor with a conventional phonon mechanism can develop an unconventional state with a nontrivial order parameter phase relation between the individual bands. We propose that such a state can explain recent experimental results on \chem{MgCNi_3}, which are suggestive of both $s$-wave pairing symmetry and of unconventional superconductivity. We show that such a state gives rise to Andreev bound states, and to spontaneous currents, at surfaces and around impurities. We also investigate possible phase transitions between states with different order parameter symmetry.
\end{abstract}


\maketitle

\newcommand{\nn}{\nonumber \\}

\newcommand{\gradient}{ {\mathbf{\nabla }} }
\newcommand{\vect}[1]{ {\mathbf{#1}} }
\newcommand{\rvec}{\vect{r}}
\newcommand{\kvec}{\vect{k}}
\newcommand{\jvec}{\vect{j}}
\newcommand{\nvec}{\vect{n}}
\newcommand{\pvec}{\vect{p}}
\newcommand{\qvec}{\vect{q}}
\newcommand{\Avec}{\vect{A}}

\newcommand{\ezmatrix}[2]{ { \left( \begin{array}{#1} #2 \end{array} \right) } }
\newcommand{\twomatrix}[4]{ \ezmatrix{cc}{ #1 & #2 \\ #3 & #4  } }
\newcommand{\twovector}[2]{ \ezmatrix{c}{ #1 \\ #2 } }
\newcommand{\threematrix}[9]{ \ezmatrix{ccc}{ #1 & #2 & #3 \\ #4 & #5 & #6 \\ #7 & #8 & #9 } }
\newcommand{\threevector}[3]{ \ezmatrix{c}{ #1 \\ #2 \\ #3 } }

\newcommand{\BdG}{Bogoliubov-de Gennes}
\newcommand{\MCN}{\chem{MgCNi_3}}
\newcommand{\Hbulk}{ \HH_{\mathrm{bulk}} }
\newcommand{\Hsurf}{ \HH_{\mathrm{surf}} }
\newcommand{\Himp}{ \HH_{\mathrm{imp}} }
\newcommand{\Heff}{ \HH_{\mathrm{eff}} }
\newcommand{\Gap}{\Delta}
\newcommand{\Pairing}{ \tilde{\Delta} }
\newcommand{\Spinor}{ \Phi }

\paragraph{Introduction.}

The discussion of unconventional superconductivity is usually limited to systems with $p$- or $d$-wave pairing. It has been noted recently that unconventional properties can also appear in systems with a multicomponent ``$s$-wave'' order parameter (OP) arising from a conventional
phonon mechanism \cite{Agterberg}. Multi-band superconductivity has been first studied shortly
after the development of the BCS theory \cite{SuhlLeggett} and has experienced a revival during recent years in connection with a variety of superconductors, including MgB$_2$, RNi$_2$B$_2$C, Sr$_2$RuO$_4$, and many others \cite{Materials}. In a multiband superconductor, in which the intraband pairing interaction is most attractive in the $s$-wave channel, each Fermi surface (FS) sheet can develop a nearly-isotropic $s$-wave OP. If the interband Cooper pair scattering is repulsive, however, a nontrivial relative OP phase configuration can exist between the bands \cite{Agterberg}. In general such a state can break symmetries in addition to $U(1)$, such as the time-reversal symmetry, responsible for unconventional behavior.

We propose here that the antiperovskite compound \cite{HeCava} \MCN\ could be a realization of this type of unconventional superconductor. The thermodynamic properties, and the observation of a Hebel-Slichter peak in NMR-measurements \cite{Singer}, suggest conventional pairing. In contrast, a zero-bias conductance peak has been observed in the quasiparticle tunneling spectra \cite{Mao}. Interpreting this as evidence for subgap Andreev bound states at the surface would point towards unconventional superconductivity \cite{Mao,Rosner}. These two seemingly contradictory observations can be reconciled within a multiband $s$-wave picture as outlined above. Our aim is to consider this type of unconventional superconductivity as a working hypothesis, and analyze observable properties that could be tested. Here we report briefly our main results; more details will be given later in a separate publication.

\begin{figure}
\includegraphics[width=7.0cm]{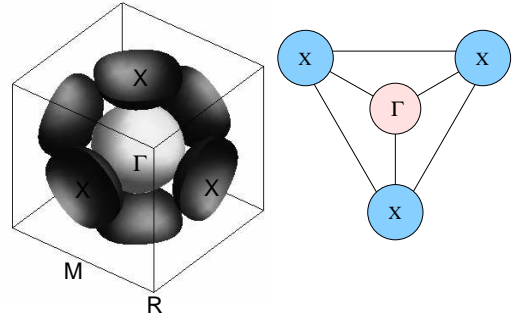}
\caption{\label{fig_bands}
Simplified view of the band structure of \MCN.}
\end{figure}

In \MCN, the important electron bands are mainly derived from a hybridization of the Ni
$3d$ with the C $2p$ orbitals. From band structure calculations \cite{BandStructure},
the main features of the FS are three pill-shaped hole pockets centered around the $X$-points, and a nearly spherical electron FS around the $\Gamma$-point, see Fig.~\ref{fig_bands}. Since these four FS sheets together comprise the largest part of the electron density of states (DOS) at the Fermi level, we ignore some minor additional sheets, namely the smaller cigar-shaped pockets along the $\Gamma$-$R$ lines, and the ``jungle gym'' structure along $M$-$R$ \cite{BandStructure}.

\paragraph{Nature of superconductivity in $\mathit{MgCNi_3}$.}

\newcommand{\eeta}{{\tilde \eta}}

Throughout this letter we maintain the viewpoint that superconductivity in \MCN\ is mainly driven by the three symmetry-related bands around the $X$-points, since their Fermi velocity is quite low in some momentum directions, resulting in a large DOS \cite{BandStructure}. These bands furthermore constitute what distinguishes \MCN\ from most ``ordinary'' BCS superconductors, and therefore are likely to contain the source of any unusual phenomena. Each of these three bands can individually develop an $s$-wave-like OP $\eta_a$, identical in magnitude by symmetry, where $a = 1,2,3$ is the band index. The repulsive Cooper pair scattering between these pockets, mediated by the Coulomb interaction, stabilizes an unconventional state satisfying $\eta_a = \eta_1 \omega^{-m(a-1)}$, where $m = \pm 1$ and $\omega = e^{2\pi i/3}$. The corresponding order parameter is $\eeta_m = \eta_1 + \omega^m \eta_2 + \omega^{-m} \eta_3$.

This can be understood if we view the Cooper pair scattering as an intrinsic Josephson coupling between different bands, which gives rise to the Josephson energy 
\begin{equation}
        E = J \left\{ \cos(\phi_{12}) + \cos(\phi_{23}) +
        \cos(\phi_{31}) \right\},
\end{equation}
where $\phi_{ab}$ is the phase difference between bands $a$ and $b$. The coupling energies are identical by symmetry. If $J$ is positive, this energy is minimized by $\phi_{12} = \phi_{23} = \phi_{31} = \pm 2\pi/3$. This two-fold degenerate state breaks time reversal symmetry. 

If the coupling between the $X$-pockets and the central $\Gamma$-band is strong enough, an unconventional state whose gap function has the $k$-space structure $\Psi_{\pm 1}(\kvec) \sim k_x^2 + \omega^{\pm 1} k_y^2 + \omega^{\mp 1} k_z^2 $ can be induced on the $\Gamma$-band as well, even if this band on its own would favor standard $s$-wave pairing. The $\eeta_{\pm 1}$ and $\Psi_{\pm 1}$ order parameters belong to the same two-dimensional representation $E_g$ of the cubic point group $O_h$, but only $\Psi_{\pm 1}$ has all the features of a ``usual'' unconventional state, namely a non-trivial phase dependence of its gap function on the FS sheet, and even nodal points at $k_x^2 = k_y^2 = k_z^2$. On the $X$-bands, the (singular) nodal points would fall in the region between the FS sheets.

\paragraph{Impurity bound states.}

\newcommand{\gcon}{g_c}
\newcommand{\gunc}{g_u}
\newcommand{\DOSX}{N_X(\epsilon_F)}
\newcommand{\DOSG}{N_{\Gamma}(\epsilon_F)}

The phase differences between the $X$-pockets, as well as the induced unconventional state on the $\Gamma$-band, can give rise to scattering-induced subgap states. Such states arise from an interference effect of Bogoliubov quasiparticles that experience a phase shift when scattered between regions of the FSs with different OP phase \cite{Tanaka}. For simplicity we will treat $X$-bands and $\Gamma$-band separately, neglecting quasiparticle scattering processes between the two. In the presence of an impurity at the origin, the $X$-bands are described by the \BdG\ (BdG) equations
\begin{equation}
\label{BdG_Impurity}
        \twomatrix{E^a_{\kvec}-\epsilon}{\Gap_a}{\Gap_a^*}{-E^a_{\kvec}-\epsilon}
        \Spinor^a_{\kvec}
        + \frac{1}{N} \sum_{\kvec' b} \twomatrix{g_{ab}}{0}{0}{-g_{ab}}
        \Spinor^b_{\kvec'} = 0,
\end{equation}
where $a,b=1\ldots 3$ label the three $X$-pockets, $E^a_{\kvec}$ denotes the kinetic energy of electrons in band $a$ with lattice momentum $\kvec$, $\Spinor^a_{\kvec} = \left( u^a_{\kvec}, \; v^a_{\kvec} \right)$ is a Nambu spinor, $g_{ab} = g_0 \delta_{ab} + g_1 (1-\delta_{ab})$ denotes intraband and interband impurity scattering matrix elements, and $N$ is the number of unit cells in the sample. In this approximation the gap functions $\Gap_a = \Gap_X \omega^{a-1}$ are spatially invariant, even in the vicinity of an impurity. Eq.~(\ref{BdG_Impurity}) is invariant under a $120^\circ$ rotation about the $[111]$-axis, since only the phase difference between the bands is physically relevant. This leads to the distinction of three sectors of eigenstates, corresponding to the three eigenvalues $1$, $\omega$, and $\omega^2$, of the rotation operator, where only the first two contain subgap states. In each sector, Eq.~(\ref{BdG_Impurity}) reduces to an effective one-band model, \eg\ for the first eigenvalue:
\begin{equation}
\label{OneBandModel}
        \twomatrix{E^1_{\kvec}-\epsilon}{\Gap_X}{\Gap_X^*}{-E^1_{\kvec}-\epsilon}
        \Spinor_{\kvec}
        + \frac{1}{N} \sum_{\kvec'} \twomatrix{g_s}{0}{0}{-g_a}
        \Spinor_{\kvec'} = 0,
\end{equation}
where the symmetrized and ``antisymmetrized'' impurity scattering matrix elements are
\begin{equation}
        g_s = \sum_a g_{ab} = g_0 + 2g_1, \;\;\;\;
        g_a = \sum_a \omega^{a-b} g_{ab} = g_0 - g_1.
\end{equation}
Note that the electron and hole components of the quasiparticles experience a different impurity scattering potential. For the second sector, $g_s$ and $g_a$ are interchanged in \Eq{OneBandModel}. It is standard to solve these equations for the bound state energies $\epsilon$, with the result 
\( 
	\label{XStates}
        \epsilon/\Delta_X = \pm \cos \left( \theta_s - \theta_a
        \right),
\)
where the scattering phase shifts are given by
\(
        \theta_{s,a} = \arctan \left[ \pi \DOSX g_{s,a} \right],
\)
$\DOSX$ being the partial density of states of a single $X$-pocket at the Fermi level. On the $\Gamma$-sheet the analogous single-band formulation gives $\epsilon/\Delta_\Gamma = \pm 1 / \sqrt{ 1 + (\pi \DOSG g_\Gamma)^2 }$, where $g_\Gamma$ corresponds to the isotropic scattering strength. Hence strong $s$-wave impurity scattering can lead to subgap bound states, with energy close to zero, for all bands.

\paragraph{Surface bound states}

\newcommand{\kin}{\kvec_{\mathrm{in}}}
\newcommand{\kout}{\kvec_{\mathrm{out}}}
\newcommand{\phin}{\phi_{\mathrm{in}}}
\newcommand{\phout}{\phi_{\mathrm{out}}}

\newcommand{\Din}{\Delta_{\text{in}}}
\newcommand{\Dout}{\Delta_{\text{out}}}
\newcommand{\phiin}{\phi_{\text{in}}}
\newcommand{\phiout}{\phi_{\text{out}}}

\newcommand{\DX}{\Delta_X}
\newcommand{\DG}{\Delta_\Gamma}

Subgap states arise also from the reflection of quasiparticles at the surface. From the BdG equations, one obtains the bound state energies (in the limit of an infinite surface barrier) 
\begin{equation}
\label{SurfaceBoundStates}
        \epsilon = \pm \frac{\mathrm{Im}\; \Din^* \Dout}
                            {|\Din - \Dout|},
\end{equation}
if the constraint $\min( |\Din|^2, |\Dout|^2 ) \ge \mathrm{Re}\; \Din^* \Dout$ is satisfied \cite{Tanaka}. If the order parameter amplitudes are equal, this expression reduces to
\(
        \epsilon
        = \pm |\Din| \cos \left[ \left( \phiin - \phiout \right) / 2 \right].
\)
Here $\Din$ and $\Dout$ denote the order parameters ``seen'' by the incoming and outgoing particle, respectively, and $\phiin$ and $\phiout$ are their respective phases. For specular scattering the momentum component parallel to the surface is conserved. At a $[100]$-surface, no subgap states appear on any band, since the phase difference is zero.  At a $[110]$-surface, the scattering processes can connect two $X$-pockets with a phase difference of $2\pi/3$, so that we expect subgap states with an energy $\epsilon \simeq \DX/2$ here. On the $\Gamma$-band, with $d$-like OP $ \Psi_{+1} $ given above, a straightforward calculation yields
\begin{equation}
\label{GammaEnergies}
        \epsilon = \pm \DG \left[ 3(k_z/k_F)^2 - 1 \right] / 2,
\end{equation}
where $\DG$ is the maximum bulk gap, and $k_z$ is the $z$-component of the quasiparticle momentum. For the $[111]$ surface, the situation is qualitatively similar on the $X$-sheets, and for the $\Gamma$-band we get
\begin{equation}
        \epsilon = \pm (\DG/2) \sin^2 \theta \cos 3 \psi,
\end{equation}
where the angles $\theta$ and $\psi$ provide a representation in spherical coordinates, so that
\begin{equation}
        \frac{\kvec}{k_F} = \frac{\cos\theta}{\sqrt{3}} \threevector{1}{1}{1}
        +\frac{\sin\theta\cos\psi}{\sqrt{2}} \threevector{1}{-1}{0}
        +\frac{\sin\theta\sin\psi}{\sqrt{6}} \threevector{1}{1}{-2}.
\end{equation}
The above bound state spectra lead to a subgap DOS
\begin{equation}
        N(\epsilon) =  \frac{\sqrt{2}A}{3}
        \left\{ \sqrt{\frac{1-\lambda}{1+2\lambda}}
         + \sqrt{\frac{1+\lambda}{1-2\lambda}} \,
             \theta{\left( \frac{1}{2} - \lambda \right)} \right\}
\end{equation}
at a $[110]$ surface, and
\begin{equation}
        N(\epsilon) = A \, \text{arcosh} \left( \frac{1}{2\lambda} \right) \,
 		\theta{\left( \frac{1}{2} - \lambda \right)}
\end{equation}
at a $[111]$ surface. Here $\lambda = \epsilon/\DG$, $A = (4/\DG) (k_F a/2\pi)^2$, and $a$ is the lattice constant. Note that in addition to a sharp peak centered at $\epsilon = 0$, a broad range of subgap states with nonzero energy is generated. 

Obviously imperfection on realistic surfaces would destroy the transverse translational symmetry and lead to diffuse scattering. Consequently, subgap states can appear even at a $[100]$ surface, and the bound state spectrum would be generally more smeared. The zero-bias features observed in quasiparticle tunneling can be interpreted as a signature of these Andreev bound states at the surface \cite{Mao}. In this way we have reconciled the observed standard form of the Hebel-Slichter peak, which results from the coherence effects of the $s$-wave pairing on each $X$-band, with the presence of subgap surface states due to the unconventional phase structure of the OP.

Additionally, we would like to emphasize that the presence of subgap states is also connected with pair breaking. In this sense, non-magnetic disorder would have an impact in this material on the the superconducting transition temperature unlike in a conventional superconductor. Interestingly, it was experimentally found that introducing C-vacancies lowers $T_c$, whereby this type of disorder is non-magnetic \cite{Doping}.

\paragraph{Ginzburg-Landau (GL) theory.}

\newcommand{\kpar}{\kvec_{||}}
\newcommand{\Dvec}{\vect{D}}
\newcommand{\Dslash}{\vect{\tilde D}}
\newcommand{\slashgrad}{\mathbf{\tilde \nabla}}

A GL description allows a qualitative analysis of spontaneous currents, which are connected with the presence of subgap states \cite{Tanaka}, and of possible phase transitions between states with different OP symmetry. In its general form, it requires three OPs $\eta_a$, $a = 1 \ldots 3$, one for each $X$-pocket, and three OPs $\Psi_0$, $\Psi_{\pm 1}$ on the $\Gamma$ sheet:
\begin{equation} 
	\left. 
	\begin{array}{l} \displaystyle
		\Psi_m \propto k_x^2 + \omega^m k_y^2 + \omega^{-m} k_z^2 \\
		\eeta_m = \eta_1 + \omega^m \eta_2 + \omega^{-m} \eta_3 
	\end{array}
	\right\} \quad \mbox{with} \quad m=0, \pm 1,
\end{equation}
where the same index $ m $ corresponds to phases of identical symmetry on the $ \Gamma $- and $X$-sheets. We focus on the $X$-bands with the $\eta$ OPs, since their superconductivity is likely strongest. For repulsive Cooper pair scattering between different $X$-pockets, $\eeta_0$ has a lower transition temperature than the degenerate $\eeta_{+1}$ and $\eeta_{-1}$ phases, and will therefore be neglected here. The GL free energy density is then
\begin{eqnarray}
	f_X &=& \sum_{m=\pm 1} \left\{ a |\eeta_m|^2 + b_1 |\eeta_m|^4
	    + K_1 |\Dvec \eeta_m|^2 \right\} \\
	  &+& b_2 |\eeta_{+1}|^2 |\eeta_{-1}|^2
	    + K_2 \left\{ (\Dvec_{+1} \eeta_{+1})^* \cdot \Dvec_{-1} \eeta_{-1} + c.c \right\},
	  \nonumber
\end{eqnarray}
where $\Dvec = (D_x, D_y, D_z) = \gradient - i \frac{2e}{\hbar c} \Avec $, and $\Avec$ is the vector potential. We also introduce the differential operators $\gradient_m = ( \partial_x, \omega^m \partial_y, \omega^{-m} \partial_z )$ and $\Dvec_m = (D_x, \omega^m D_y, \omega^{-m} D_z)$. The current density is then
\begin{equation}
        \jvec = -\frac{4e}{\hbar c}  \sum_{m = \pm 1}
		\text{Im} \left\{ K_1 \; \eeta_m^* \gradient \eeta_m
		  + K_2 \; \eeta_m^* \gradient_m \eeta_{-m}
        \right\}.
\end{equation}

\paragraph{Currents around an impurity.}

\newcommand{\qslash}{\tilde q}
\newcommand{\du}{\delta \eeta_{+1}}
\newcommand{\dv}{\delta \eeta_{-1}}

If the impurity does not break any point group symmetries, it will introduce a term
\(
        f_{\text{imp}} =  S \left( |\eeta_{+1}|^2 + |\eeta_{-1}|^2 \right)
        \delta^3(\rvec)
\)
in the free energy. In an $ \eeta_{+1} $ bulk state also an $\eeta_{-1}$ component will be induced around this defect. Denoting the deviations of  $\eeta_{+1}$ and $\eeta_{-1}$ from their bulk value by $\delta \eeta_{\pm}$, we arrive at the GL equations (to first order in S) 
\begin{eqnarray}
        \left[ 2a + K_1 \gradient^2 \right] \du + K_2 \gradient_{-1}^2 \, \dv 
                   &=& S \eeta_{+1}^{\text{bulk}} \, \delta^3(\rvec), \nn
        \left[ 2b + K_1 \gradient^2 \right] \dv + K_2 \gradient_{+1}^2 \, \du &=& 0,
\end{eqnarray}
where $2b = -a-b_2 |\eeta_{+1}^{\text{bulk}}|^2$. These equations can be solved in Fourier space, giving rise to the currents
\begin{equation}
        \jvec_q = -\frac{4 i e}{\hbar c}
        \frac{S |\eeta_{+1}^{\text{bulk}}|^2 K_2^2 \; \text{Im}\left[ Q^* \tilde \qvec \right]}
        {(2a-K_1 q^2)(2b-K_1 q^2) - K_2 |Q|^2},
\end{equation}
where $\tilde \qvec = (q_x, \omega q_y, \omega^2 q_z)$ and $Q = q_x^2 + \omega q_y^2 + \omega^2 q_z^2$. This corresponds to a angular structure
\begin{equation}
        \jvec(\rvec) \sim (x(y^2-z^2),y(z^2-x^2), z(x^2-y^2))/r^3,
\end{equation}
which is an octupole. Consequently, the resulting magnetic field has a basic quadrupolar structure with corrections of angular momentum $l=4$, i.e.\ the field vanishes directly at the impurity site.

\paragraph{Surface currents.}

\newcommand{\omegaslash}{\tilde \omega}
\newcommand{\nslash}{\tilde \nvec}
\newcommand{\xx}{x'}

At a boundary or interface, the following surface terms enter the GL expansion:
\begin{equation}
        f_{\text{surf}} = \delta(\xx) \left[ S_1 ( |\eeta_+|^2 + |\eeta_-|^2 )
        + S_2 ( \eeta_+^* \omegaslash \eeta_- + c.c. ) \right],
\end{equation}
where $\xx = \nvec \cdot \rvec$, $\omegaslash = n_x^2 + \omega n_y^2 + \omega^2 n_z^2$, and $\nvec = (n_x, n_y, n_z)$ is the surface normal. A calculation similar to that for the impurity current yields a surface current
\begin{equation}
        \jvec \sim |\eeta_{+1}^{\text{bulk}}|^2 \;
        \text{Im}\left[ \omegaslash^* \nslash \right],
\end{equation}
which is localized on a length scale comparable to the correlation length. Here $\nslash = (n_x, \omega n_y, \omega^2 n_z)$. A spontaneous current is generated at a $[110]$ surface, where $\text{Im} \left[ \omegaslash^* \nslash \right] \sim  (-1, 1, 0)$, but not at $[100]$ or $[111]$ surfaces, where $\text{Im} \left[ \omegaslash^* \nslash \right]$ is zero.

\paragraph{Competition between $d$- and $s$-wave OPs.}

$\Psi_{\pm 1}$ is an externally induced OP that competes with the ``natural'' $\Psi_0$ OP on the $\Gamma$ band, which has a different ($s$-wave like) symmetry than the inducing source. If the gap introduced by $ \Psi_{\pm 1} $ is large enough, it will govern the quasiparticles on the $ \Gamma $-Fermi surface entirely and be the only OP of this band. Weakening this induced pairing state, however, can open the door for a state with a finite $ \Psi_0 $-component. Therefore, at $ T = 0 $ a quantum phase transition is possible for the $ \Gamma $-band as a function of the impurity concentration. To illustrate this at least on a qualitative level, we consider the GL theory describing this competition,
\begin{eqnarray}
        f_{\Gamma} &=& \sum_{m=0,+1} \left\{ a'_m |\Psi_m|^2 + b'_m |\Psi_m|^4
        \right\} + \gamma |\Psi_{+1}|^2 |\Psi_0|^2 \nn
		& & + g ( \Psi^*_{+1} \eeta_{+1} + c.c. ),
\end{eqnarray}
where we regard $\eeta_{+1}$ as an external OP that is assumed to be constant, and ignore any feedback of the $ \Gamma $-band on the $ X$-bands. For $\eeta_{+1} = 0$, only the $s$-wave state $ \Psi_0 $ is stable, and the coexistence of $\Psi_{+1}$ and $\Psi_0$ is energetically disfavored. Hence $a'_{+1}$ and $\gamma$ are positive, while $a'_0$ is negative at temperatures below $T_{c0}$ ($<T_c$), the transition temperature of $\Psi_0$. The phase of $\Psi_0$ is determined by additional terms involving $\Psi_{-1}$, which were omitted in the above expression. These are small, since $\Psi_{-1}$ is suppressed by the coupling to the $X$-bands, so that the phase of $\Psi_0$ is only weakly pinned. Collective phase modes may hence be relevant near the transition \cite{KumarWoelfle}.

\begin{figure}
\includegraphics[width=6.5cm]{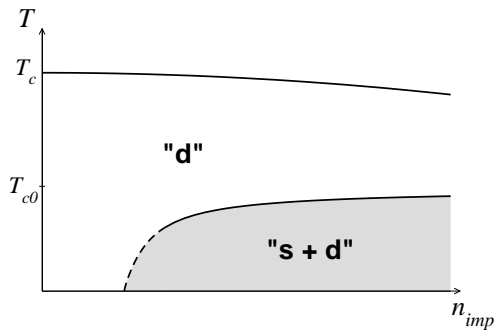}
\caption{\label{fig_phasediagram}
Schematic phase diagram for the
$\Psi_{+1}$ and $\Psi_0$ order parameters on the central band, in the 
temperature vs.\ disorder plane.
Solid lines mark second-order
phase transitions; the dashed line marks a region in which the transition
may be first-order if the coupling between $\Psi_{+1}$ and $\Psi_0$ is strong
enough.
}
\end{figure}

We may influence the strength of the $\Psi_{+1}$-component by non-magnetic impurities which with increasing concentration suppress this unconventional pairing channel of the $\Gamma$-band, while the ``$s$-wave'' state is basically unaffected. The growing concentration of impurities, $n_{\text{imp}}$, leads to the increase of $a'_{+1}$, causing the decrease of $\Psi_{+1}$. A continuous onset of $\Psi_0$ occurs when $a'_0 + \gamma | \Psi_{+1}|^2 = 0$ is satisfied. However, if the coupling coefficient $\gamma$, i.e.\ the repulsion between the two OPs, is sufficiently strong, then a first order transition is favored within this model, see Fig.~\ref{fig_phasediagram}. In both cases the new phase appearing corresponds to the coexistence of the two OP, since $\Psi_{+1}$ enforced from ``outside'' can never be eliminated completely.

\paragraph{Conclusions.}

In summary, we have shown that the --- at first glance contradictory --- experimental results on \MCN\ (Hebel-Slichter peak in NMR and zero-bias anomaly in tunneling spectra) can be reconciled by a multi-band unconventional superconducting state, where the OP on the dominant FS sheets is $s$-wave like, but with non-trivial phase relations between the bands. The resulting state violates time reversal symmetry. Subgap Andreev bound states and spontaneous currents are signatures of such a state that are accessible to experiment. Our specific predictions about the direction dependence of surface bound state energies can be tested in tunneling experiments on single crystals. The spontaneous impurity-induced currents can be verified by zero-field muon relaxation measurements \cite{Luke}, while the spontaneous surface currents might be observed by scanning SQUID or Hall probes. Furthermore, the induced superconducting state on the central $\Gamma$-band can undergo a transition from a ``$d$-wave'' state to an ``$s + d $''-wave state with increasing impurity concentration. The first- or second-order transition (see Fig.~\ref{fig_phasediagram}) may or may not occur, depending on the precise values of microscopic parameters. Such a transition may leave their signature in thermodynamic quantities such as specific heat, London penetration depth, or ultrasound absorption.  Eventually we believe that MgCNi$_3$ could be the exciting case of an unconventional superconducting state due to its multi-band structure \cite{Agterberg}. 

We would like to thank D.F.~Agterberg, Ch.~Helm, H.~Kusunose, and T.M.~Rice for helpful discussions. This work was financially supported by the Swiss National Science Foundation.

\newcommand{\journal}[4]{#1~\textbf{#2}, #3 (#4)}
\newcommand{\nature}[3]{\journal{Nature}{#1}{#2}{#3}}
\newcommand{\PRL}[3]{\journal{Phys.\ Rev.\ Lett.}{#1}{#2}{#3}}
\newcommand{\PRB}[3]{\journal{Phys.\ Rev.\ B}{#1}{#2}{#3}}
\newcommand{\RMP}[3]{\journal{Rev.\ Mod.\ Phys.}{#1}{#2}{#3}}
\newcommand{\JPC}[3]{\journal{J.\ Phys.\ C}{#1}{#2}{#3}}
\newcommand{\JPSJ}[3]{\journal{J.\ Phys.\ Soc.\ Jpn.}{#1}{#2}{#3}}
\newcommand{\JETP}[3]{\journal{JETP}{#1}{#2}{#3}}
\newcommand{\XJETP}[6]{\journal{Zh.\ Eksp.\ Teor.\ Fiz.}{#1}{#2}{#3}
  [\JETP{#4}{#5}{#6}]}
\newcommand{\PMB}[3]{\journal{Phil.\ Mag.\ B}{#1}{#2}{#3}}
\newcommand{\EPL}[3]{\journal{Europhys.\ Lett.}{#1}{#2}{#3}}

\newcommand{\JAppPhys}[3]{\journal{J.\ App.\ Phys.}{#1}{#2}{#3}}
\newcommand{\PhysicaC}[3]{\journal{Physica C}{#1}{#2}{#3}}
\newcommand{\SolStat}[3]{\journal{Sol.\ Stat.\ Comm.}{#1}{#2}{#3}}
\newcommand{\ProgTheorPhys}[3]{\journal{Prog.\ Theor.\ Phys.}{#1}{#2}{#3}}
\newcommand{\RepProgPhys}[3]{\journal{Rep.\ Prog.\ Phys.}{#1}{#2}{#3}}

\newcommand{\condmat}[1]{preprint: cond-mat/#1}

\end{document}